\documentclass[manuscript,screen]{acmart}

\AtBeginDocument{%
  \providecommand\BibTeX{{%
    \normalfont B\kern-0.5em{\scshape i\kern-0.25em b}\kern-0.8em\TeX}}}

\copyrightyear{2026}
\acmYear{2026}
\setcopyright{cc}
\setcctype{by}
\acmConference[AutomotiveUI '26]{18th International Conference on Automotive User Interfaces and Interactive Vehicular Applications}{September 20--23, 2026}{Gothenburg, Sweden}
\acmBooktitle{18th International Conference on Automotive User Interfaces and Interactive Vehicular Applications (AutomotiveUI '26), September 20--23, 2026, Gothenburg, Sweden}
\acmDOI{10.1145/3828157.3828774}
\acmISBN{979-8-4007-2814-3/2026/09}

\usepackage{subcaption} % captions for subfigures and the like

\usepackage{url}
\makeatletter
\g@addto@macro{\UrlBreaks}{\UrlOrds}
\makeatother

\usepackage{multirow} %tabular cells spanning multiple rows

\usepackage{color} %colour management

\usepackage{enumitem} 

\usepackage[official]{eurosym}

\usepackage{xspace}

\usepackage{xcolor, colortbl}% http://ctan.org/pkg/xcolor

\usepackage{arydshln}

\newcommand{\m}{\textit{M=}}

\newcommand{\sd}{\textit{SD=}}

\newcommand{\N}{\textit{N=}}

\newcommand{\F}[3]{$F({#1},{#2})={#3}$}

\newcommand{\p}{\textit{p=}}

\newcommand{\padj}{\textit{p$_{adj}$=}}

\newcommand{\pminor}{\textit{p$<$}}

\newcommand{\infoContent}{\textit{information content}\xspace}

\newcommand{\VIP}{\textit{sightedness}\xspace}

\def\plaintitle{Effects of Auditory Information for People With Visual Impairments in Highly Automated Vehicles}

\begin{document}

%[short-title]
\title[Effects of Auditory Information for BVI in Highly Automated Vehicles]{\plaintitle}

\author{Mark Colley}
%\authornote{Both authors contributed equally to this research.}
\email{m.colley@ucl.ac.uk}
\orcid{0000-0001-5207-5029}
\affiliation{%
  \institution{UCL Interaction Centre}
  \city{London}
  \country{United Kingdom}
}

\author{Tobias Aescht}
\email{tobias.aescht@uni-ulm.de}
\orcid{0000-0002-3055-8289}
\affiliation{%
  \institution{Ulm University}
  \city{Ulm}
  \country{Germany}
}

\author{Omid Rajabi}
\email{omid.rajabi@uni-ulm.de}
\orcid{0000-0003-4498-2539}
\affiliation{%
  \institution{Institute of Media Informatics, Ulm University}
  \city{Ulm}
  \country{Germany}
}

\author{Max R{\"a}dler}
\email{max.raedler@uni-ulm.de}
\orcid{0000-0002-5413-2637}
\affiliation{%
  \institution{Ulm University}
  \city{Ulm}
  \country{Germany}
}

\author{Pascal Jansen}
\email{pascal.jansen@uni-ulm.de}
\orcid{0000-0002-9335-5462}
\affiliation{%
  \institution{Ulm University}
  \city{Ulm}
  \country{Germany}
}

\author{Enrico Rukzio}
%\authornotemark[1]
\email{enrico.rukzio@uni-ulm.de}
\orcid{0000-0002-4213-2226}
\affiliation{%
  \institution{Ulm University}
  \city{Ulm}
  \country{Germany}
}
% 150 words or less
% What is the large scope and problem space? Why should we care? Motivation

% What is the specific problem addressed? Problem

% Why is the problem Important? Why was this work carried out?

% What have you done? Solution

% What did you find out? What are the concrete results?

% What are the implications on a larger scale? How does it change the bigger picture?

\renewcommand{\shortauthors}{Colley et al.}

% Do not change the page size or page settings.
\begin{abstract}
Automated vehicles promise to improve accessibility and access to personal mobility for everyone. However, their design and current research trends in visualizing relevant information do not reflect this commitment to accessibility for users with visual impairments.
Therefore, we designed and implemented a visual and auditory communication concept for people with visual impairments seated inside fully automated vehicles. 
Furthermore, in an online video-based study (N=35, 12 with visual impairments), we compared three levels of auditory information communication: low (safety-relevant information only), medium (additionally including vehicle control and route updates), and high (additionally including sightseeing and destination information).
Results showed that trust and user experience significantly improved with additional information, with a corresponding, albeit less robust, effect on perceived safety. However, they also revealed that a potential information saturation was reached with medium information. 
Our work helps to improve the accessibility of automated vehicles by guiding designers towards adequate information communication.
\end{abstract}



\maketitle


\section{Introduction}
%Motivation
Worldwide, there are around 1.3 billion people with some vision impairment, 217 million people with moderate to severe vision impairment, and 36 million people who are blind~\cite{bourne2017magnitude, who_impaired}. Nonetheless, in the USA, for example, 80\% of the legally blind population still have useful residual vision~\cite{civilRightsUSAblind}. Until now, people who are blind and visually impaired (BVI) could not benefit from individual mobility, as they could not operate vehicles or require special production~\cite{sucu2013haptic}. Lack of (individual) mobility reduces the ability to socialize, access health care, go shopping, and even gain employment~\cite{stenquist2014self, bennett2019attitudes}. However, with the advent of automated vehicles (AVs), this could change~\cite{brinkley_international_2019, brinkley_exploring_2020, fink_fully_2021}. 

%Problem
With such novel technology, challenges with over-and undertrust emerge. With undertrust, the use of this technology could be scarce, and thus the potential improvements in overall traffic and personal mobility could not become a reality. In surveys of the general population, \citet{kyriakidis2015public} found that 65\% of the participants were worried about the reliability of automated cars. Continental AG found that between 43 and 74\% of the survey participants doubted whether AVs would function reliably~\cite{continentalMobility2013}. \citet{schoettle2014survey} report that 75\% were at least slightly worried about system failure in unforeseen situations. \citet{brinkley_international_2019} also states that trust towards AVs is currently low for BVIs. 
These worries are likely even higher for people with disabilities, as this group can not intervene in the case of an emergency.

Prior work has explored a range of visualization techniques (mostly shown on dashboard and windshield displays) to improve or calibrate trust in AVs, including highlighting vehicles in adverse weather~\cite{wintersberger2019fostering}, indicating pedestrian intent~\cite{colley2020effect}, visualizing detected objects~\cite{colley2021effects}, presenting predicted trajectories~\cite{schneider2021explain, colley2024effects}, and comparing alternatives in critical scenarios~\cite{colley2021should}. These approaches rely on visual encodings as their primary communication channel. 
However, such approaches can fail for BVIs. Visualizations may be partially or fully imperceptible, difficult to interpret due to reduced visual acuity or contrast sensitivity, or cognitively inaccessible when critical information is encoded in spatial or temporal visual patterns. As a result, the underlying system state and intent can remain opaque for BVIs. Yet, research explicitly addressing their perceptual and interaction requirements remains limited, and many AV technologies under development are not accessible to BVIs~\cite{brinkley_exploring_2020}.

%Solution
Therefore, we designed and implemented a prototypical visual information communication design within the AV dashboard to leverage the remaining vision of BVIs, for example, by employing high contrast and highly visible colors. One of the authors has a visual disability (right eye: visual acuity of 0.1; left eye: blind with light perception) and thus was able to assess prototypes early in the design phase.

However, our main focus and evaluation were on the auditory design. Audition is, besides touch, the primary non-visual channel through which BVIs perceive their environment~\cite{Chanana.2017}, and speech-based output is the interaction modality most frequently requested by BVI users in the context of AVs~\cite{brinkley_exploring_2020, brewer2020supporting}. Unlike tactile displays, auditory output requires no additional hardware and no learned encoding. At the same time, auditory information is transient, competes with conversation and environmental sounds, and risks annoyance when overused. This trade-off raises the question of \textit{how much} information an AV should communicate auditorily---the question at the core of this work. The information communication included safety-relevant (e.g., start/stop, warning signals), vehicle control, route updates, sightseeing, and destination-related information.

In an online video-based study, both with sighted people (\textit{n=}23) and BVIs (\textit{n=}12), we evaluated three different auditory information communication conditions: with low (only safety-relevant information), medium (additionally including vehicle control and route updates), and high (compared to medium, additionally including sightseeing and destination) information (see \autoref{tab:informationcategoriesAndConditions}). The dashboard's visual design remained consistent across all conditions. 
We found that trust, understanding, and user experience improved for sighted participants and BVIs when comparing the low \infoContent to medium and high auditory information. However, effects were not uniform, with BVIs often preferring the medium \infoContent condition. 

\textit{Contribution Statement:} The contributions of this work are (1) a visual design for information communication of driving-relevant information tailored to the needs of BVIs. (2) An auditory communication design with different levels of auditory information. (3) Results of an online video-based study with \N{35} participants (12 BVIs). % shows that medium information mainly improves perceived safety and reduces cognitive load.

\section{Related Work}
Our work builds on previous focus groups and studies with BVIs on AV accessibility. We also briefly highlight related work on visualizing relevant information about trust increases and calibration. 

\subsection{Including People with Visual Impairments in the Design of Automated Vehicles}
In BVI focus groups, some respondents expressed a need for in-vehicle feedback to improve situation awareness (SA), particularly regarding location-related information. BVIs also indicated some concern about the human-machine interface (HMI) regarding the interaction of the BVI with the AV in terms of information transfer and vehicle operation~\cite{brinkley_opinions_2017}.

\citet{brinkley_international_2019} proposed a system for the interaction of BVIs with AVs called ATLAS (short for \textit{Accessible Technology Leveraged for Autonomous Vehicles System}). ATLAS allows control of the AV (e.g., setting a destination) by regulating user input and providing feedback. The system includes features like auditory voice feedback while driving to meet situation awareness (SA) needs. The system also uses spatial audio and visual technologies, such as displays, to meet location verification needs. Information presentation such as direction, traffic information, or points of interest (POI) is triggered by geofences~\cite{hickeys_guidelines_nodate}. Results showed that users answered positively about whether ATLAS increased their SA. %This already shows good results using only auditory feedback in this area. However, efforts can be made to improve this further by using additional methods or by making this system less interruptive, so that, for example, conversations between two vehicle occupants are not interrupted by a ``talking in between'' vehicle.

Other works~\cite{brinkley_exploring_2020, brewer_understanding_2018,pakusch_unintended_2020, brewer2020supporting} addressed how AVs can communicate information to visually impaired users. Still, no previous work has specifically addressed the question of \textbf{what} the information needs of BVIs are while driving in AVs and \textbf{how} they can be met.
However, work by \citet{brinkley_exploring_2020} with focus groups indicated that 53\% of participants expressed a desire for a location verification system for relevant environmental objects, and that such a system was important to some degree. While this partly addresses \textbf{what} information is required, it provides no guidance on \textbf{how} this need can be met. Likewise, 45\% of the participants indicated that a system to increase the passenger's environmental awareness was needed (see also~\cite{brinkley_opinions_2017}). This system should provide a real-time representation of the AV's relationship to other relevant objects and road users, such as vehicles, buildings, and pedestrians, mirroring the information a sighted driver would have through cameras, windows, and mirrors. Also, 71\% of participants stated that they would prefer a voice-based system to interact with the vehicle~\cite{brinkley_exploring_2020}.
Additionally, it was stated that an HMI for a blind or low-vision user must meet the needs for SA, location verification, and other needs not yet specified due to a lack of research~\cite{brinkley_exploring_2020}. Subjects also expressed a desire for a fully AV to have a system to verify arrival at the correct destination~\cite{brinkley_exploring_2020}. Finally, concerns were expressed about traveling without directional information and current surroundings, such as the AV going in the wrong direction or to the wrong destination without being noticed in the vehicle.

\citet{brewer_understanding_2019} interviewed BVIs about their experiences with ride-sharing services such as Uber or Lyft, showing that these BVI passengers like to receive information about the environment (e.g., local landmarks being passed or doors and obstacles at the destination) from drivers during the journey. If the drivers of these services did not communicate such information, the BVI passengers indicated they would ask for it. In their overview, \citet{brewer2020supporting} also highlight that primarily speech-based systems are appropriate to provide information about their surroundings to increase their SA. Nonetheless, different degrees of disabilities should be supported. Therefore, participants suggested multiple modalities, including tactile systems such as a tactile compass or a vibration system, to present obstacles and provide environmental awareness~\cite{brewer2020supporting}.

\citet{lee_eliciting_2021} show that the characteristics of different populations, such as people with disabilities or children, must be considered in the design of AV applications and interfaces. These groups have been denied the benefits of personal mobility until now; therefore, their needs must be taken into account toward a more inclusive future mobility. Furthermore, the work revealed that passengers should be provided with journey information to improve their SA. Although passengers no longer have to drive themselves, participants emphasized the importance of information such as destination, travel time, vehicle environment, and the vehicle's intentions~\cite{lee_eliciting_2021}.

\subsection{Auditory Displays in Vehicles and for People with Visual Impairments}\label{sec:rw_auditory}
Auditory cues have repeatedly been shown to support awareness in traffic contexts. \citet{Schoop.2018} demonstrated that cyclists' awareness of nearby vehicles could be significantly enhanced using auditory cues varying in direction, tempo, pitch, and timbre, mapped to the direction, distance, and type of detected vehicles. In the context of AVs, \citet{Gang.2018} improved passenger SA by incorporating earcons within a 3D sound environment. \citet{nadri2021novel} compared auditory and visual feedback and their combination, finding that audio-visual feedback notably increased participants' SA compared to a visual-only representation, and highlighting the differential roles of verbal audio and auditory icons. Relatedly, \citet{Glatz.2018} found that auditory icons are particularly effective for conveying contextual information, while verbal audio is better suited for delivering time-critical information. For BVIs specifically, \citet{Brinkley.2019} developed a prototype that enhances SA within AVs via audible location cues and spatial audio, increasing trust, SA, and perceived reliability.

Taken together, prior work establishes that auditory and audio-visual feedback can improve SA and trust in vehicle contexts and that speech and non-speech audio serve complementary roles. However, these studies primarily varied the \textit{modality} of feedback. \textit{Which} information, and \textit{how much} of it, should be communicated auditorily to BVI passengers of AVs has not been systematically investigated; our study addresses this gap.

\subsection{Visualizing General Information About Automated Vehicle Capabilities}
Previous work evaluated communicating the AV's detection of other vehicles~\cite{wintersberger2019fostering, colley2021effects, lindemann2018catch}, pedestrians~\cite{colley2021effects, colley2020effect, colley2021should, wilbrink2020reflecting, lindemann2018catch} and their intentions~\cite{colley2020effect, colley2022effectsScene}. Besides visualizing which road users were detected, the intention of the ego vehicle was visualized, for example, via LED strips~\cite{locken2016autoambicar} or showing the trajectories on the windshield~\cite{colley2024effects}. 
Highlighting the capabilities of an AV was proposed to improve trust (e.g.,~\cite{wintersberger2019fostering}); however, visualized uncertainty, either in the driving task as a whole~\cite{beller2013improving,helldin2013presenting} or in AV's object detection capabilities~\cite{colley2021effects, colley2022effectsScene}, was proposed to calibrate trust.
\citet{kunze2018augmented} employed simulated Augmented Reality (AR) to display uncertainties of longitudinal and lateral control (i.e., the ability to steer and accelerate/decelerate). 
While these works argue for the usage of AR (e.g., \cite{colley2021effects, colley2020effect, colley2022effectsScene}) and their visualizations are partially transparent, we argue that when including BVI, visibility must be the highest priority. Therefore, while this work focuses on auditory information, the displays inside the AV should have high contrast.

\section{Experiment}

We designed and conducted a mixed-design online video study with \N{35} (12 of whom were BVI) to evaluate the effects of different feedback strategies. \infoContent was manipulated within-subjects with three levels (low vs. medium vs. high; see \autoref{tab:informationcategoriesAndConditions}): every participant experienced all three videos. \VIP (BVI vs. sighted) was a quasi-experimental between-subjects factor.
This study was guided by the exploratory research question:

\noindent\textit{What impact does the level of auditory information detail (\infoContent; within-subjects) have on sighted and visually impaired AV users (\VIP; between-subjects) in terms of (1) cognitive load, (2) situation awareness, (3) trust, (4) perceived safety, and (5) user experience?}

%\todo{wenig Info über auditory design generell}
\subsection{Design of the Auditory Information Communication}

%\citet{Schoop.2018} demonstrated that cyclists' awareness of nearby vehicles could be significantly enhanced using auditory cues that varied in direction, tempo, pitch, and timbre. These auditory cues were mapped to indicate the direction, distance, and vehicle type detected, providing cyclists with intuitive, immediate information. In the context of AVs, \citet{Gang.2018} applied this modality to improve passenger SA by incorporating earcons within a 3D sound environment. Similarly, \citet{nadri2021novel} conducted a comparative study on the effectiveness of auditory versus visual feedback and their combination. Audio-visual feedback notably increased participants' SA compared to a visual-only representation. This study also highlighted the differential impact of verbal audio and auditory icons, in line with \citet{Glatz.2018}, who found that auditory icons are particularly effective for conveying contextual information, while verbal audio is better suited for delivering time-critical information. For BVIs, \citet{Brinkley.2019} developed a prototype system that enhances their SA within AVs by utilizing audible location cues and spatial audio. Their research showed that this system increased trust and SA and enhanced perceived reliability.\citet{Chanana.2017} stated that the auditory channel, besides the tactile one, is among the most used for pedestrians. Therefore, we focused on this modality.

For the auditory design, a combination of sounds and speech was used. Sounds were used as these are easily distinguishable and can communicate different meanings without becoming annoying~\cite{walker2013spearcons, nadri2021novel}.
The audio files used for the sounds were retrieved from \url{notificationsounds.com}~\cite{noauthor_ringtones_nodate} under the Creative Commons Attribution 4.0 International Public License. For a description of the sound, please see Appendix. For the start sound, we used the sound ''Insight''\footnote{Insight: \url{https://notificationsounds.com/sound-effects/insight-578}; accessed March 31, 2026}, for the end sound ''Done for you''\footnote{Done for you: \url{https://notificationsounds.com/notification-sounds/done-for-you-612}; accessed March 31, 2026}, for crossing pedestrians ''I demand attention''\footnote{I demand attention: \url{https://notificationsounds.com/message-tones/i-demand-attention-244}; accessed March 31, 2026} and for driving maneuvers (left and right turn, roundabout) the sound ''Gesture''\footnote{Gesture: \url{https://notificationsounds.com/message-tones/gesture-192}; accessed March 31, 2026}. 

The Google Cloud Text-To-Speech Engine (Wavenet-G)~\cite{noauthor_text--speech_nodate} was used for the AV's voice output.
We distinguished five levels of information based on the work of \citet{meinhardt2024hey}: Safety relevant, Vehicle control, Route updates, Sightseeing, and Destination (see \autoref{tab:informationcategoriesAndConditions} left). Sightseeing was added as an opportunity for AVs to support BVIs in engaging with their surroundings.

The assignment of the five categories to the three \infoContent conditions (\autoref{tab:informationcategoriesAndConditions}, right) follows a cumulative nesting by functional criticality. Safety-relevant information (A) constitutes the minimal baseline: it concerns events of immediate physical relevance and corresponds to the warning function that vehicles provide, irrespective of accessibility considerations. The medium condition adds operational transparency---what the vehicle is doing (B) and how the journey is progressing (C)---which prior work has consistently identified as the core information need of BVI passengers~\cite{brinkley_opinions_2017, brinkley_exploring_2020, brewer_understanding_2019, lee_eliciting_2021}. The high condition further adds discretionary, context-enriching information: engagement with the surroundings (D) and verification of the destination environment (E), needs voiced in prior focus groups~\cite{brinkley_exploring_2020, brewer_understanding_2019} but not required for a safe and comprehensible ride. We chose cumulative (nested) rather than crossed combinations for two reasons: nesting guarantees a strict ordering of conditions by information volume---a prerequisite for investigating saturation effects---whereas crossing all categories ($2^5$ combinations) would have confounded information \textit{amount} with information \textit{type} and exceeded an acceptable study duration.

\begin{table*}[ht!]
\footnotesize
\begin{tabular}{ ll }  

  \begin{tabular}{ll|l}
    \toprule
    Level & Information type & Content \\
    \midrule
     A & Safety relevant & indications, warning signals, start/stop  \\
     B & Vehicle control & steering, vehicle intentions, journey start/end \\
     C & Route updates & progress, time change, section \\
     D & Sightseeing & Sights, POIs, Environment \\
     E & Destination & environment information, navigation aids \\
  \bottomrule
\end{tabular}

& 

  \begin{tabular}{l|l}
    \toprule
    \infoContent & Information levels conveyed \\
    \midrule
    low information & A \\
    medium information &  A, B, C \\
    high information &   A, B, C, D, E\\
  \bottomrule
\end{tabular}

\end{tabular}

\caption{Information categories (left) and conditions used in the experiment (right) according to the defined information categories.}
\label{tab:informationcategoriesAndConditions}
\end{table*}

\begin{table*}[ht!]
\scriptsize
\centering
\caption{Classified notifications. The number in the brackets refers to the occurrence order in the condition ``High information''.}
  \begin{tabular}{p{2.3cm}|lp{11.0cm}}
    \toprule
    Event &  Level & Text (if used) \\
    \midrule
    Start &  A & Sound ''Insight'' \\
    Stopping &  A & Sound ''Done for you'' \\
    Sudden stop for pedestrians & \hspace{0.1cm} B & Sound ''I demand attention'' \\
    Turning, going through roundabout & \hspace{0.1cm} B &  (8) ``We are now stopping at the A66 highway entry.''; \newline (9) ``We are now entering the A360 highway.''; \newline (11) ``We are now leaving the A360 highway, via the exit A67 rest area Koenigsberg.''; \newline (11) ``We have reached our destination the Koenigsberg rest area. It is 10:46 am.''\\
    Traffic light, crosswalk & \hspace{0.1cm}  B & Sound ''Gesture'' \\
    Route information & \hspace{0.2cm} C & (1) ``You have chosen the Koenigsberg rest area as your destination. The estimated travel time is 5 minutes. Our estimated time of arrival is 10:45 a.m.''\\
    Route updates/section & \hspace{0.2cm} C &  (2) ``Our next section of the journey is in the urban area, so there may be frequent stops at pedestrian crossings and traffic lights. The expected traffic volume is moderate.''; \newline (6) ``“We are now leaving the city of Koenigsberg. Our next section of the journey is through a rural area, which may lead to uneven roads. The expected volume of traffic is low.''; \newline (10) ``We are now leaving the rural area. Our next drive section is on the A360 highway. The expected volume of traffic is high.''\\
    Event/route change (road works/accident)  & \hspace{0.2cm} C & (4) ``There is an unforeseen road closure due to a construction site in front of us. We will take an alternative route, which will delay our expected arrival by 1 minute to 10:46 a.m.''; \newline (7) ``There is an unforeseen road closure due to an accident in front of us. The emergency services are already on site. We now stop at the accident scene and wait for the oncoming lane to clear so we can continue our journey.''; \\
    Information about POIs  & \hspace{0.3cm} D & (3, not in medium) ``We will now pass the famous Albert Einstein statue on the left''; \newline (5, not in medium) ``In front of us is now the statue of the famous war horse of Koenigsberg.''\\
    %Information about the current environment & \hspace{0.9cm} D & \\
    %Information about other road users  & \hspace{0.9cm} D & \\
    Information about the destination  & \hspace{0.4cm} E &  (12, not in medium) To our right is a meadow, in front of us is a parked truck, and behind us, there is another car. There is also a garbage can on the right, and on the street to our right are benches. The area to the left of us is free to get off to the street, it is also possible to get off to the left on the meadow'' \\
  \bottomrule
\end{tabular}
\label{tab:informationcategoriesspecific}
\end{table*}

\autoref{tab:informationcategoriesspecific} shows all the provided information in the AV course.
While comparing all information levels (and their combinations) could have yielded more specific insights, we implemented three levels of \infoContent (see \autoref{tab:informationcategoriesAndConditions} on the right): low, medium, and high. This helped us to keep the study in an acceptable time frame.

\subsection{Design of the Driving Environment and Automated Vehicle Information Dashboard}\label{sec:materials}

We designed a journey through a city, followed by a drive through a rural area and a short highway section in Unity~\cite{unitygameengine}. %We modeled the scenario in Unity version 2020.3.3f~\cite{unitygameengine}. 

\begin{figure}[htb]
\centering
\includegraphics[width=0.37\linewidth]{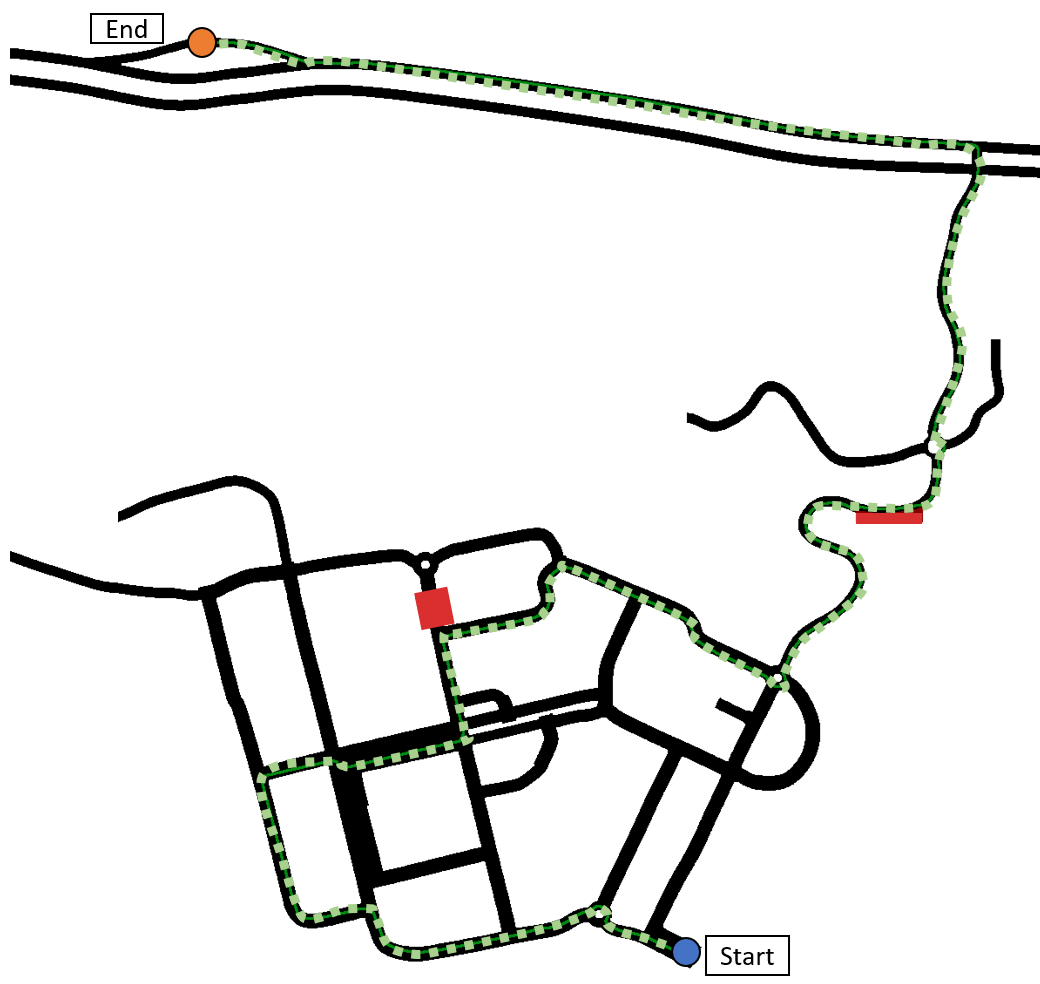}
\caption{Route through the fictitious city of Königsberg (in green), unexpected obstacles on the route (in red).}\label{fig:map}
\Description{Route through the fictitious city of Königsberg. The route features numerous curves, two roundabouts, and some signalized intersections. There are parts in the city and parts on the highway.}
\end{figure}

\begin{figure*}[ht!]
    \begin{subfigure}[c]{0.37\textwidth}
        \centering
            \includegraphics[width=\textwidth]{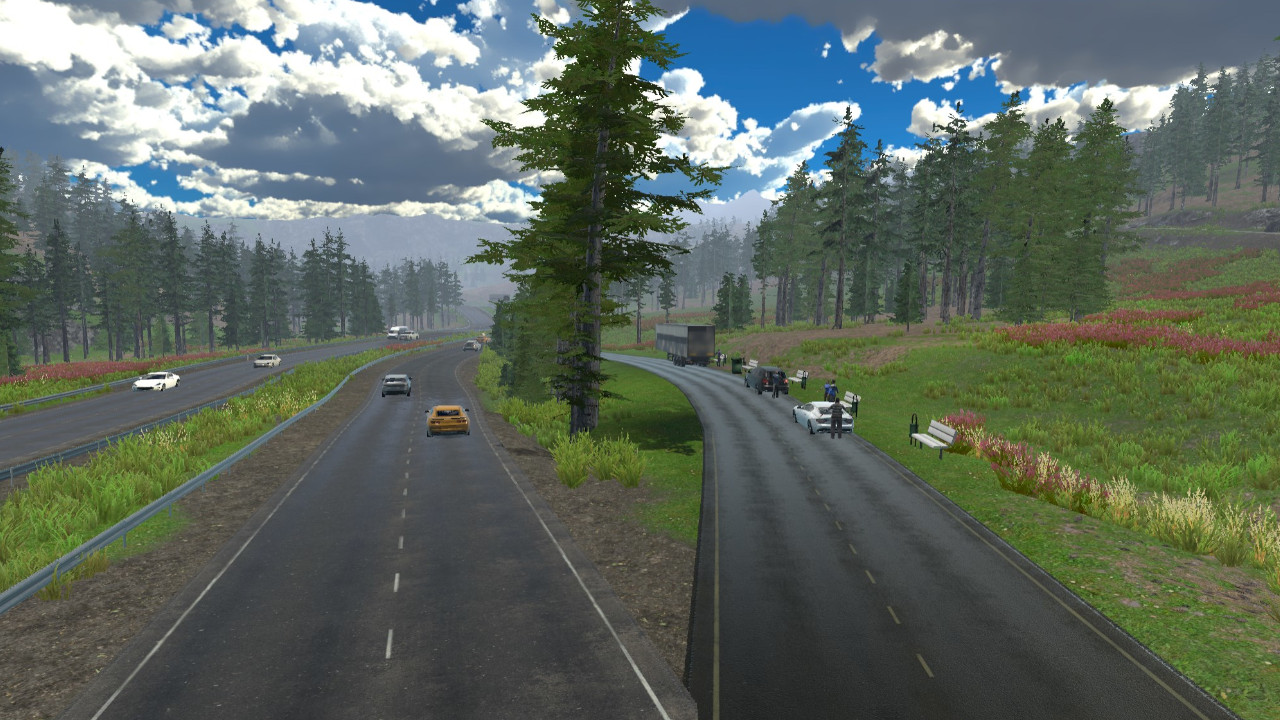}
        \subcaption{Final destination: rest area.}\label{fig:rest_area}
    \end{subfigure}
    \begin{subfigure}[c]{0.37\textwidth}
        \centering
            \includegraphics[width=\textwidth]{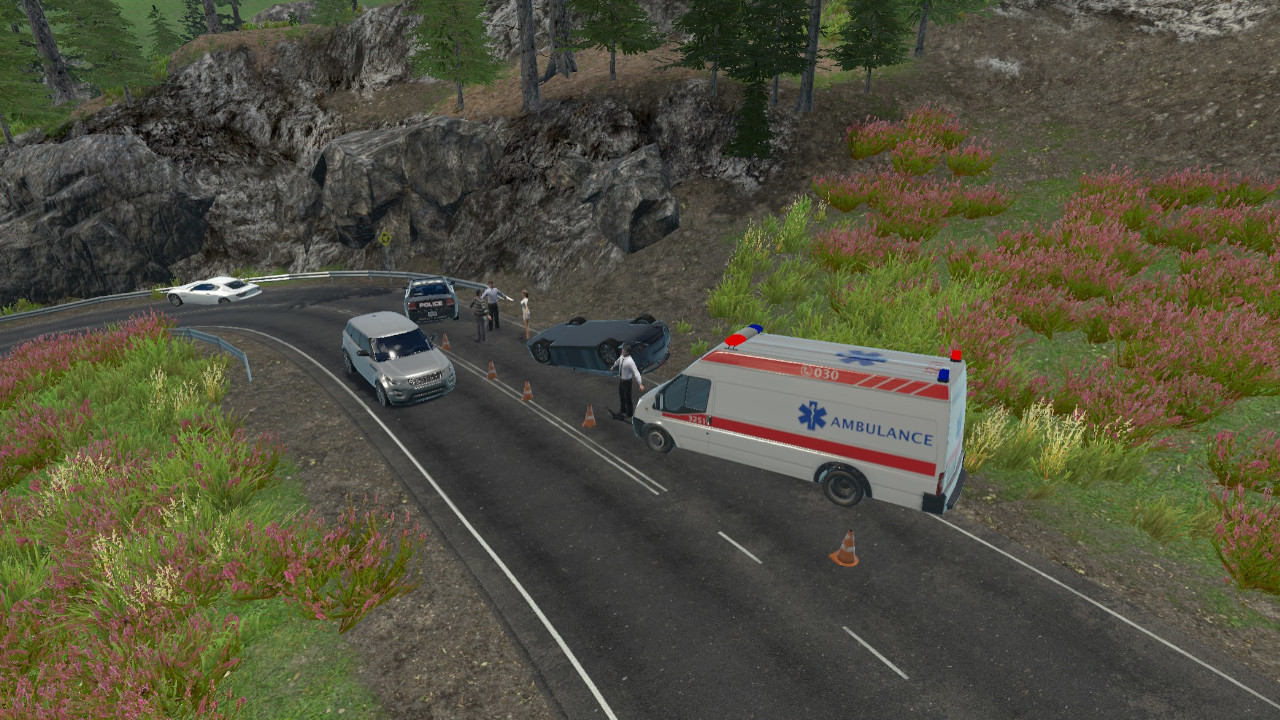}
        \subcaption{Accident site.}\label{fig:unfall}
    \end{subfigure}
   \caption{Rest area and the accident site.}
   \Description{Subfigure a shows the final destination, a rest area. On the left, the highway continues in both directions. In subfigure b, an accident occurs. A car is turned over, and an ambulance stands in front of it, blocking the road.}
\end{figure*}

\begin{figure*}[htb]
\centering
\includegraphics[width=0.7\textwidth]{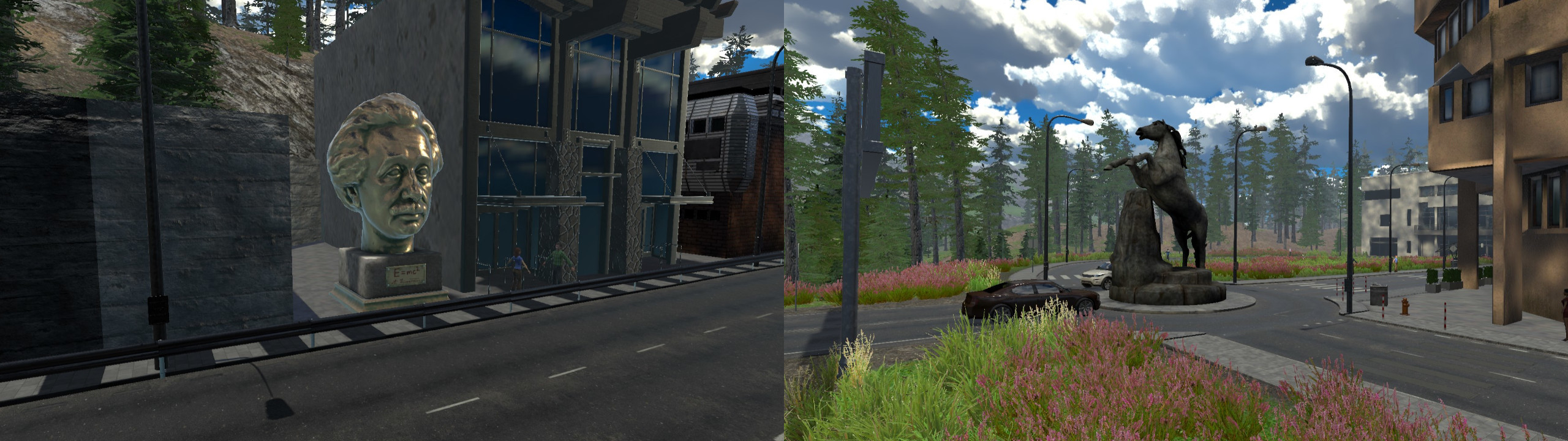}
\caption{POIs. Left: Albert Einstein statue, Right: Quarrel of Königsberg.}\label{fig:unity:pois}
\Description{This figure is divided into two subfigures. On the left, a bronze/golden statue of Albert Einstein stands between two buildings. On the right, there is a statue in the middle of a roundabout called the Quarrel of Königsberg.}
\end{figure*}

\begin{figure*}[htb]
\centering
\includegraphics[width=0.6\textwidth]{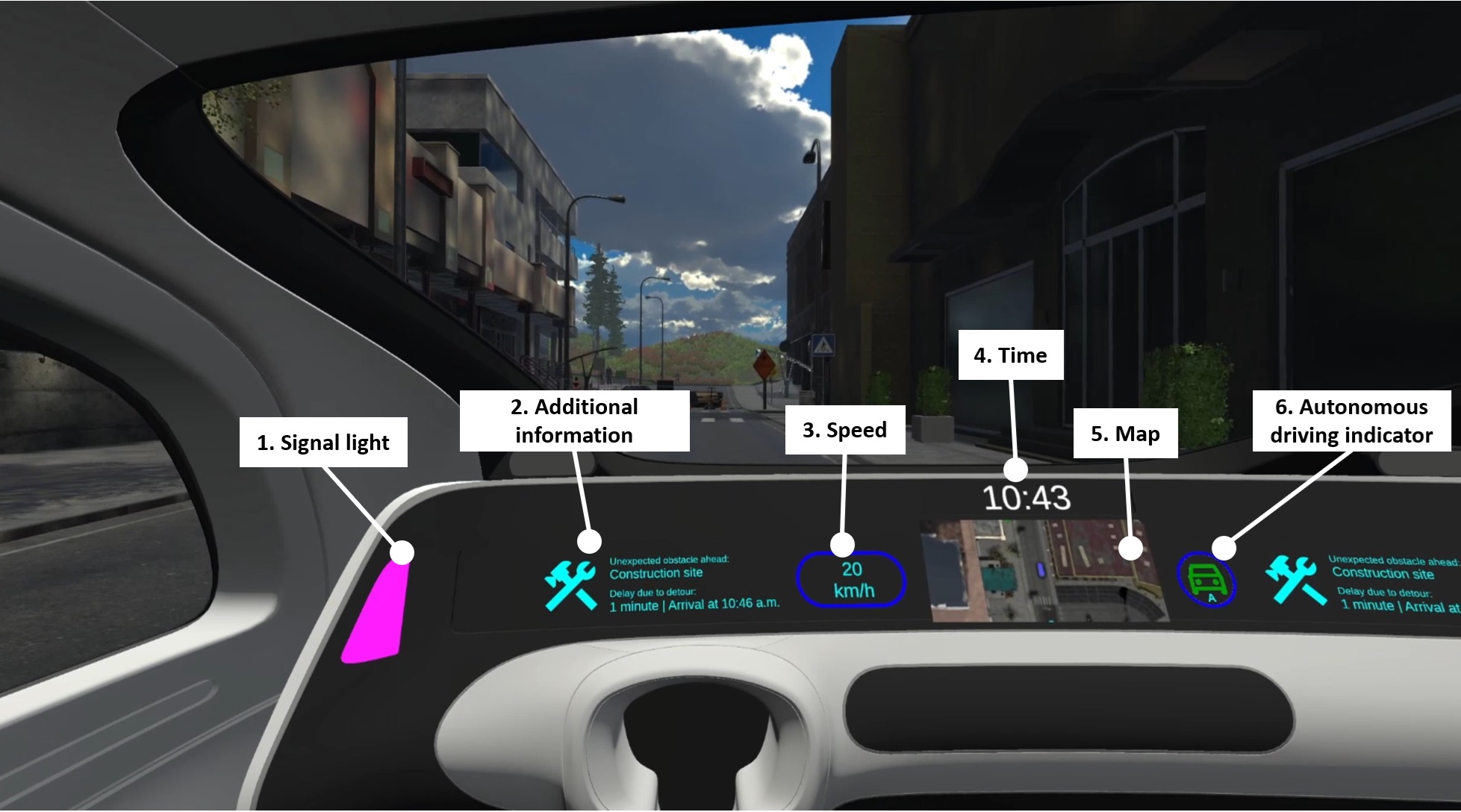}
\caption{Dashboard of the AV. 1. Colored signal lights indicating normal driving-related actions of the vehicle via cyan and magenta for unexpected events, 2. information displayed on the dashboard, with large icons and text, 3. current speed, 4. time, 5. map showing surroundings, 6. autonomous driving indicator.}\label{fig:unity:dashboard}
\Description{This figure shows the dashboard of the vehicle. The relevant areas are described via text.}
\end{figure*}

%\subsubsection{Environment and Visual Design}
%Windridge City~\cite{noauthor_windridge_nodate} was used as the environment. 
For simulating vehicles and other people, a modified and extended version of the Urban Traffic System~\cite{noauthor_urban_nodate} was used. Windridge City~\cite{noauthor_windridge_nodate} was extended to include environmental elements for the route, including a rest area to serve as a destination outside of the city which was created using EasyRoads3D~\cite{noauthor_easyroads3d_nodate} (see \autoref{fig:rest_area}), a construction site (see \autoref{fig:unity:dashboard} in the background) through which the route must be unexpectedly adjusted in the simulation and an accident site (see \autoref{fig:unfall}) as an unexpected obstacle which the AV must communicate and overcome.
We also added two landmarks (see \autoref{fig:unity:pois}) as POIs to provide environment-specific information. Parked vehicles and passers-by were added to create a lively city environment. 

Due to the route, there are numerous driving-related actions, such as turning or driving through a traffic circle (see \autoref{fig:map}). Likewise, street crossings for passers-by, both at designated crossings and at unexpected locations in the inner-city area of the route, were taken from the original route guidance and adapted. A Mercedes-Benz F015 concept vehicle~\cite{noauthor_mercedes-benz_nodate} served as AV.
The steering wheel was removed for the simulation to represent that it is an AV, and thus, intervention in driving is impossible.

%\subsubsection{Automated Vehicle Information Dashboard Design}
\begin{figure*}[htb]
\centering
\includegraphics[width=0.9\textwidth]{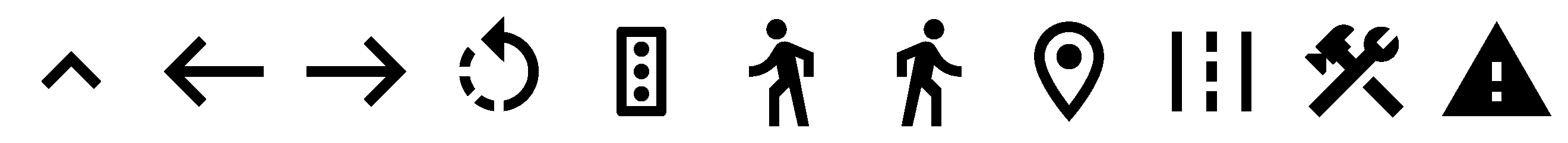}
\caption{Icons used in the visual design. Left to right: start, turn left, turn right, traffic circle, traffic light, passer-by from right, passer-by from left, place, routes section, construction site, and unexpected obstacle. Taken from Google Material Icons~\cite{noauthor_google_nodate}.}\label{fig:icons}
\Description{This figure shows different icons in black.}
\end{figure*}

The visual design followed established guidance for low-vision accessibility. Because most legally blind people retain usable residual vision~\cite{civilRightsUSAblind}, interfaces should maximize luminance contrast, use large sans-serif type, avoid conveying meaning through color alone, and minimize animation and visual clutter~\cite{etsi_normbarrierefrei_nodate, wcag21}. We applied these principles to a dashboard configuration that otherwise reflects information likely to be available in AVs designed for a general audience.

A dashboard was added to the AV interior (see \autoref{fig:unity:dashboard}). This dashboard represents one of many possible configurations. The dashboard is only marginally adjusted to the needs of BVIs, as we believe that future AVs will be designed mostly for able-bodied people. However, as with smartphones, there will likely be some BVI-related adjustment possibilities. Therefore, we adjusted a potential interface to ensure proper visibility, achieved through sufficient size and contrast, as confirmed by the author with a visual disability.
Additionally, the visualized information should not cause mental overload but still serve as a reasonable information source. Therefore, no animations and a reduced information load were visualized. Nonetheless, we included factors likely to be available in AVs determined for the general audience. The dashboard displays an icon for each action on the left and, on the right, additional text describing the current action or situation (see \autoref{fig:icons}). The following describes the AV dashboard design.

%\todo{change red to less visible in figure}
In the center of the dashboard, a map of the vehicle's surroundings shows passers-by as green dots, vehicles as rectangles, other vehicles in cyan, and the ego vehicle in dark blue. This additional highlighting, with clearly visible colors, helps gather the relevant information quickly. This helps to get an overview of the environment detected by the vehicle. This resembles the visualizations used by \citet{colley2020effect}. To the left of the map is the AV's current speed. Above the map, there is the time of day. An indicator that the vehicle is operating in fully autonomous mode is displayed to the right of the map. This is currently available information that we believe will also be relevant in the future for a general audience (not only BVI users; e.g., see the necessity of automation mode marker in the context of external communication of AVs~\cite{faas2021self}).

The colored-light system displays different AV actions in the user's periphery, in addition to textual information via color (see \autoref{fig:unity:dashboard} left). After internal tests, we used only two colors while driving to keep the system's complexity low. The colors cyan and magenta were chosen because they are highly conspicuous compared to other colors used in road traffic and are not associated with meanings such as red for warning or danger~\cite{werner_new_2018}. The lights illuminate in cyan for normal vehicle driving-related actions, such as turning, going through a traffic circle, or stopping at a traffic light. This indicates that a vehicle is constantly moving on this route. In the case of unexpected actions or events, such as pedestrians suddenly crossing the road or a spontaneous route closure due to a construction site or an accident, the lights are magenta, signaling a situation that could not have been foreseen but is handled by the vehicle without issue. At the beginning and end of the journey, the lights are white to signal that the vehicle has switched from active driving mode to parking or standby mode, or vice versa. 
Visualization of uncertainty, as, for example, argued for by \citet{colley2021effects}, is not a viable option for people with disabilities, as they are not capable nor are they allowed to take over control.

\begin{comment}
\begin{table}[ht!]
\footnotesize
\centering
  \begin{tabular}{l|l}
    \toprule
    \infoContent & Information levels conveyed \\
    \midrule
    low information & A \\
    medium information &  A, B, C \\
    high information &   A, B, C, D, E\\
  \bottomrule
\end{tabular}
\caption{Conditions used in the experiment, information levels after \autoref{tab:informationcategories}.}
\label{tab:conditions}
\end{table}
\end{comment}

\subsection{Measurements}\label{subsec:measurements}

We employed the cognitive load subscale of the raw NASA-TLX~\cite{hart1988development} on a 20-point scale (``How much mental and perceptual activity was required? Was the task easy or demanding, simple or complex?''; 1=Very Low to 20=Very High). 
Additionally, we used the subscale \textit{Trust} of the \textit{Trust in Automation} questionnaire by \citet{korber2018theoretical}. Trust is measured via agreement using 5-point Likert scales (1=\textit{Strongly disagree} to 5=\textit{Strongly agree}) on two statements (``I trust the system.'' and ``I can rely on the system.''). While additional subfactors are relevant~\cite{kaur2018trust}, we focused on a short questionnaire to avoid overlong questionnaires for BVI (as done in several automotive-related questionnaires~\cite{colley2020towards, colley2020evaluating, manchon2021manual, colley2022effects, colley2021investigating, hartwich2021improving}). Subjective SA was measured using the 10-dimensional Situation Awareness Rating Technique (SART)~\cite{taylor2017situational} with its subscales Demand, Supply, and Understanding. Regarding the user experience, the short version of the User Experience Questionnaire (UEQ-S)~\cite{schrepp2017design} with the subscales pragmatic and hedonic quality was used.
Perceived safety was rated using four 7-point semantic differentials from -3 (anxious/agitated/unsafe/timid) to +3 (relaxed/calm/safe/confident)~\cite{faas2020longitudinal}. Participants also rated their agreement with the statement ``All relevant information has been communicated.'' on a 5-point Likert scale (1=\textit{Strongly disagree} to 5=\textit{Strongly agree}). 

After all three videos (i.e., one per \infoContent), participants rated the information's reasonability and necessity on a 7-point Likert scale. They also rated the different information types (Start/End information, Driving related information, Pedestrian crossing information, Route information, Information about unforeseen events (i.e., accident or construction site), Sightseeing \& POI information (i.e., statues), Information about surroundings) from 1=\textit{not useful at all} to 5=\textit{extremely useful}. Participants also rated the general design of the in-vehicle information systems (1=\textit{Very bad} to 5=\textit{Very good}) regarding the \textit{Light} (colored lights located left and right of the dashboard), \textit{Information displayed} on the dashboard screen, \textit{Sound} signals, and the \textit{Voice} feedback.
Finally, participants provided open feedback.

\subsection{Procedure}

Each session started with a brief introduction, followed by an audio check, a demographic questionnaire, and a signature on the consent form. Then, participants were instructed to use a laptop or a computer.
We introduced participants as follows: 
%\begin{quote}
\noindent\textit{In this study, you will see 3 videos from the ego perspective of a person riding in a fully autonomous vehicle. The car has a built-in information system that delivers information to the driver in various ways. In the 3 videos, the car will drive the same route through a city and its outskirts. The information provided will differ from video to video.
While watching the videos, imagine you are experiencing the scenes unfold before you.}
%\end{quote}

Then, participants were introduced to the dashboard's visual design, as shown in \autoref{fig:unity:dashboard}, and to the sounds. Participants were required to indicate their familiarity with the dashboard design and the sound.

After the introduction, the participants were assigned to the three conditions in counter-balanced order (balanced Latin square). After the video, participants answered the questionnaires detailed in \autoref{subsec:measurements}. Every video took approximately six minutes.

A script ran in the background, keeping the window maximized and preventing participants from skipping or rerunning the video (to ensure equal exposure time). A Full HD (1920 × 1080) monitor and loudspeakers were required. The questionnaire could be used with screen readers for blind participants. On average, a session lasted $\approx$35 min. Participants were compensated with \EUR{3.50}.
The experimental procedure followed the ethics committee guidelines of our university and adhered to regulations on the handling of sensitive and private data, anonymization, compensation, and risk aversion. Compliant with our university‘s local regulations, no additional formal ethics approval was required.

\section{Results}

\subsection{Data Analysis}
Before every statistical test, we checked required assumptions (normality and homogeneity of variance). 
For non-parametric data, we used aligned rank transform (ART) via the ARTool package~\cite{wobbrock2011art} and Holm correction for post-hoc tests. For parametric data, we used an ANOVA.
For \autoref{fig:reasonable}, we used the R package \texttt{ggstatsplot} in version 1.0.0~\cite{ggstatsplot}. These include the mean or median, density plots, boxplots, and data points.
They also include statistical details (test used, number of observations, effect size, confidence interval). Therefore, we do not rewrite these in text.
R in version 4.6.0 and RStudio in version 2026.05.1 were employed. All packages were up to date in June 2026.

\subsection{Participants}
We determined the required sample size via an a priori power analysis using G*Power v3.1.9.7~\cite{faul2009statistical}. To achieve a power of .8 with an alpha level of .05, 28 participants should result in an anticipated medium to high effect size (0.25~\cite{funder2019}) in a within-between-subjects design with two groups and three measurements.

%\todo{intention for occupying the users?}

\N{35} people took part in the study (ten female, 25 male). The mean age of the participants was \m{33.46} years (range = 18 to 74, \sd{13.75}). Twelve participants (34\%) reported visual impairments. Three of them (25\%) reported having a (self-defined) moderate visual impairment (visual acuity worse than 6/60), three (25\%) reported having a severe visual impairment (visual acuity worse than 3/60), two (17\%) reported having an extreme visual impairment, and three (25\%) reported being blind. One person did not provide further information on their visual impairment.  Participants indicated on a 5-point Likert scale that there was general interest in autonomous driving (\m{4.03}, \sd{1.27}) and that they expected autonomous driving could make their lives easier (\m{3.97}, \sd{1.22}). Also, the statement ''I think autonomous driving will become a reality in the next 10 years. (by 2031).'' was more likely to be agreed with (\m{3.86}, \sd{1.14}). 
However, these questions already showed differences (albeit not significant) between the two groups. In all 3 items, BVIs responded more positively, so there was generally a higher interest in autonomous driving (BVI: \m{4.25}, \sd{1.22}; sighted: \m{3.91}, \sd{1.31}; W = 164.5, p = 0.33), also expectations that autonomous driving would simplify their lives were higher (BVI: \m{4.42}, \sd{1.00}; sighted: \m{3.74}, \sd{1.29}; W = 184, p = 0.09). Expectations that autonomous driving will become a reality in the next 10 years were also higher for the BVI group (BVI: \m{4.25}, \sd{0.97}; sighted: \m{3.65}, \sd{1.19}; W = 179.5, p-value = 0.15).

\subsection{Cognitive Load, Trust, and Perceived Safety}
The ART found a significant main effect of \VIP on cognitive load (\F{1}{33}{4.64}, \p{0.039}, $\eta_{p}^{2}$ = 0.12, 95\% CI: [0.00, 1.00]). Sighted participants reported higher (\m{10.33}, \sd{5.40}) cognitive load  than BVIs (\m{7.11}, \sd{6.18}). We found no significant interaction effects.

% ATTENTION: Difference between NPAV and ART
%The NPAV found a significant interaction effect of \infoContent $\times$ \VIP on cognitive load  (\F{2}{66}{3.48}, \p{0.036}; see~\autoref{fig:tlx_ie}). While cognitive load decreased with increasing auditory information content for sighted participants, only the medium information content reduced cognitive load for BVI participants.

%\begin{figure}[ht!]
%    \centering
%    \includegraphics[width=0.45\textwidth]{figures/results/tlx_interaction.pdf}
%    \caption{Interaction effect on cognitive load .}
%     \Description{This figure shows the interaction effect on cognitive load. For sighted participants, cognitive load declined with additional information; VIP reported the lowest cognitive load at the medium information level.}
%    \label{fig:tlx_ie}
%\end{figure}

The ART found a significant main effect of \infoContent on trust (\F{2}{66}{9.94}, \pminor{0.001}, $\eta_{p}^{2}$ = 0.23, 95\% CI: [0.09, 1.00]). Post-hoc tests found the differences between the low information condition (\m{3.10}, \sd{1.33}) compared to medium (\m{3.97}, \sd{0.81}, \padj{0.017}) and high (\m{3.94}, \sd{0.91}, \padj{0.013}) to be significant. Trust was lowest in the low information condition. We found no significant interaction effects.

The ART found a significant main effect of \infoContent on perceived safety (\F{2}{66}{3.56}, \p{0.034}, $\eta_{p}^{2}$ = 0.10, 95\% CI: [0.00, 1.00]). However, post-hoc tests found no significant differences. We found no significant interaction effects.

%The NPAV found a significant interaction effect of \infoContent $\times$ \VIP on perceived safety (\F{2}{66}{4.35}, \p{0.017}; see \autoref{fig:ps_ie}). While perceived safety increased with information content for both groups, the increase from low to medium information was higher for BVI. Additionally, while perceived safety remained approximately equal across the medium and high information conditions for sighted participants, it decreased from medium to high for BVI participants.

% \begin{figure}[ht!]
%     \centering
%     \includegraphics[width=0.45\textwidth]{figures/results/ps_score_interaction.pdf}
%     \caption{Interaction effect on perceived safety.}
%      \Description{This figure shows the interaction effect on perceived safety. Perceived safety was lowest for sighted participants with little information, but higher and almost equal for medium and high information. For VIP, perceived safety was clearly highest for the medium information.}
%     \label{fig:ps_ie}
% \end{figure}

\subsection{Situation Awareness}
A mixed ANOVA found a significant main effect of \infoContent on SA (\F{2}{66}{4.57}, \p{0.014}, $\eta_{p}^{2}$ = 0.12, 95\% CI: [0.02, 1.00]). Post-hoc tests showed these differences were not significant (low: \m{16.49}, \sd{10.34}; medium: \m{19.69}, \sd{5.60}; high: \m{21.37}, \sd{7.46}). 
Regarding the subscales, the ART found a significant main effect of \infoContent on Demand (\F{2}{66}{3.84}, \p{0.027}, $\eta_{p}^{2}$ = 0.10, 95\% CI: [0.01, 1.00]). However, post-hoc tests found no significant differences (low: \m{13.74}, \sd{4.79}; medium: \m{12.51}, \sd{3.64}; high: \m{11.71}, \sd{4.33}). 
A mixed ANOVA also found no significant effect on Supply (low: \m{18.00}, \sd{5.69}; medium: \m{17.20}, \sd{4.73}; high: \m{17.31}, \sd{4.90}). 
The ART found a significant main effect of \infoContent on the Understanding subscale (\F{2}{66}{7.90}, \pminor{0.001}, $\eta_{p}^{2}$ = 0.19, 95\% CI: [0.06, 1.00]). Post-hoc tests showed the difference between low (\m{12.23}, \sd{5.67}) and high information (\m{15.77}, \sd{3.67}, \padj{0.026}) to be significant.
We found no significant interaction effects on any of the subscales nor SA.

\subsection{User Experience and Presented Information}
% ATTENTION: DIFFERENCE to NPAV
%The NPAV found a significant main effect of \infoContent on pragmatic quality (\F{2}{66}{9.03}, \pminor{0.001}). Post-hoc tests showed that the differences between the low information condition (\m{4.66}, \sd{1.82}) and medium (\m{6.24}, \sd{0.76}) and high (\m{5.99}, \sd{0.99}) information conditions were significant.

\begin{figure}[ht!]
    \centering
    \begin{minipage}[t]{0.49\textwidth}
        \centering
        \includegraphics[width=0.75\textwidth]{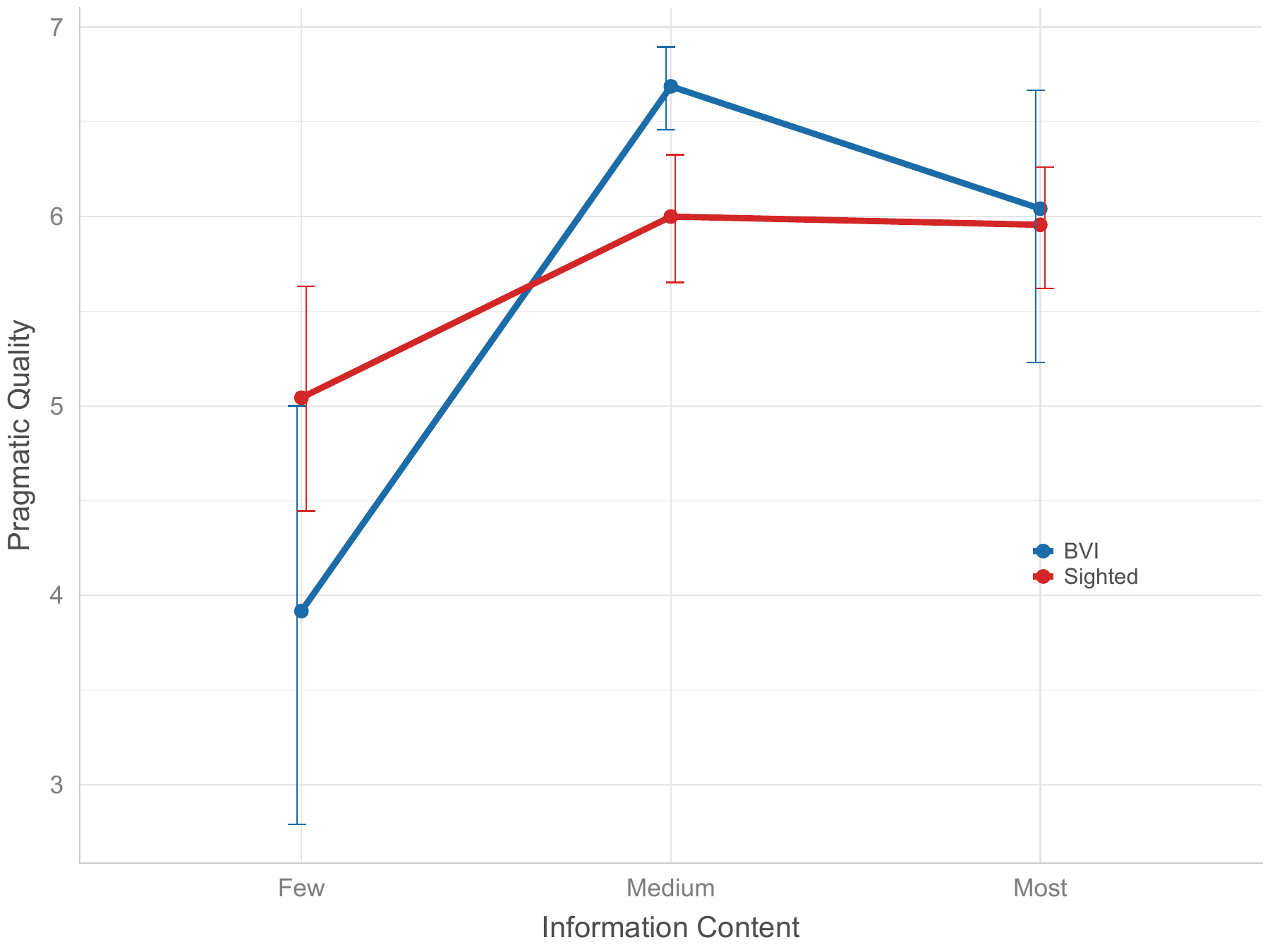}
        \caption{Interaction effect on pragmatic quality.}
        \Description{This figure shows the interaction effect on pragmatic quality. Pragmatic quality was lowest for sighted participants with little information, but higher and almost equal for medium and high information. For BVI, pragmatic quality was clearly highest for the medium information.}
        \label{fig:pq_ie}
    \end{minipage}
    \hfill
    \begin{minipage}[t]{0.49\textwidth}
        \centering
        \includegraphics[width=0.75\textwidth]{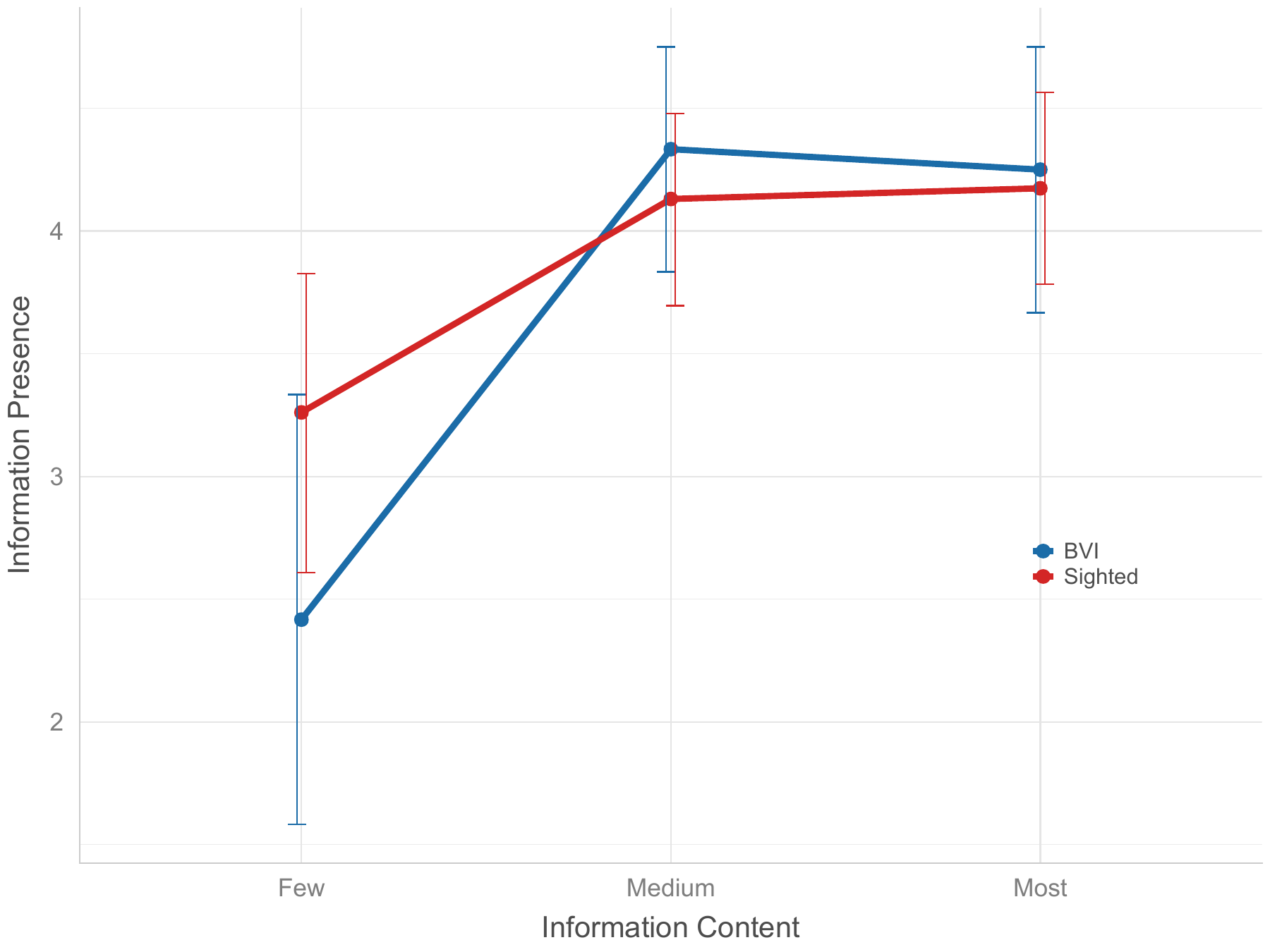}
        \caption{Interaction effect on \textit{presence of all relevant information}.}
        \Description{This figure shows the interaction effect on the presence of all relevant information. While the presence of all relevant information increased with information content for both groups, the increase from low to medium information was higher for BVI. Additionally, the agreement on the presence of all relevant information remained approximately equal across sighted participants and BVIs in the medium and high information conditions.}
        \label{fig:presence_ie}
    \end{minipage}
\end{figure}

The ART found a significant main effect of \infoContent (\F{2}{66}{17.73}, \pminor{0.001}, $\eta_{p}^{2}$ = 0.35, 95\% CI: [0.19, 1.00]) and a significant interaction effect of \infoContent $\times$ \VIP on pragmatic quality (\F{2}{66}{6.68}, \p{0.002}, $\eta_{p}^{2}$ = 0.17, 95\% CI: [0.04, 1.00]; see \autoref{fig:pq_ie}). While pragmatic quality increased with information content for both groups, the increase from low to medium information was higher for BVI. Additionally, while the pragmatic quality remained approximately equal across the medium and high information conditions for sighted participants, it decreased from medium to the high information condition for BVI participants.

The ART found a significant main effect of \infoContent on hedonic quality (\F{2}{66}{12.32}, \pminor{0.001}, $\eta_{p}^{2}$ = 0.27, 95\% CI: [0.12, 1.00]). 
Post-hoc tests showed the differences between the low (\m{4.14}, \sd{2.19}) and medium (\m{5.53}, \sd{1.48}, \padj{0.020}) and high (\m{5.56}, \sd{1.31}, \padj{0.014}) condition to be significant. Hedonic quality was lowest in the low information condition. We found no significant interaction effects.

% ATTENTION: DIFFERENCE TO NPAV
%The NPAV found a significant main effect of \infoContent on presence of all relevant information (\F{2}{66}{12.80}, \pminor{0.001}). Post-hoc tests showed that the differences between the low information condition (\m{2.97}, \sd{1.54}) and medium (\m{4.20}, \sd{0.90}) and high (\m{4.20}, \sd{0.93}) information were significant.

The ART found a significant main effect of \infoContent (\F{2}{66}{22.44}, \pminor{0.001}, $\eta_{p}^{2}$ = 0.40, 95\% CI: [0.25, 1.00]) and a significant interaction effect of \infoContent $\times$ \VIP on \textit{presence of all relevant information} (\F{2}{66}{6.09}, \p{0.004}, $\eta_{p}^{2}$ = 0.16, 95\% CI: [0.03, 1.00]; see \autoref{fig:presence_ie}). While the \textit{presence of all relevant information} increased with information content for both groups, the increase from low to medium information was higher for BVI. Additionally, the agreement on the \textit{presence of all relevant information} remained approximately equal for sighted participants and BVIs across the medium and high information conditions.

\subsection{Final Questions and Open Feedback}
\begin{figure}[ht!]
    \centering
    \includegraphics[width=0.38\textwidth]{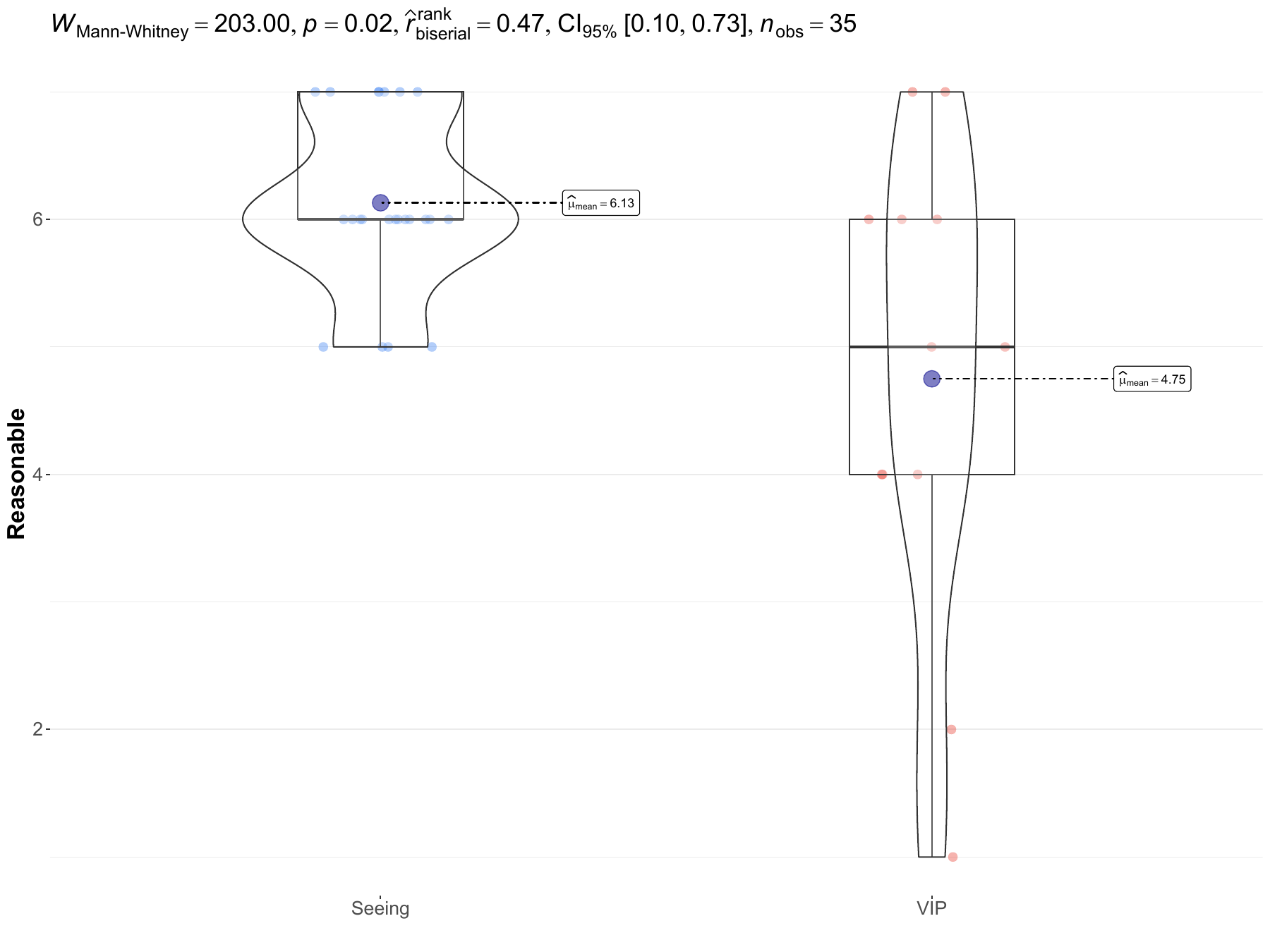}
    \caption{Significant difference for reasonability ratings. Sighted participants found the reasonability to be higher. The high variance for BVI should be highlighted.}
     \Description{Box and density plots of reasonability ratings for BVI and sighted participants. Sighted participants' ratings cluster at the upper end of the scale with little spread, while BVI participants' ratings span nearly the entire scale, indicating large between-person variance.}
    \label{fig:reasonable}
\end{figure}
A Mann-Whitney U test found a significant difference between BVI and sighted participants regarding the reasonability of the provided information (see \autoref{fig:reasonable}). The information was rated significantly more reasonable by the sighted participants. However, there was a large variance for the BVI in reasonability assessments.

Mann-Whitney U tests found no significant difference between the BVI and sighted participants in terms of necessity (\p{0.445}) or usage (``I would use the provided information while riding autonomous vehicles.'') of the provided information (\p{0.462}). The results for the other dependent variables measured after the three videos were shown are listed in~\autoref{tab:ratings}.

\begin{table*}[ht!]
\centering
\footnotesize
\caption{Values and statistical comparison (Mann-Whitney U test) between BVI and sighted participants after the three videos.}
\begin{tabular}{@{}lllll@{}}
\toprule
                            & \textbf{BVI}        & \textbf{sighted}     & \textbf{Mann-Whitney U test}   \\ \midrule
\textit{usefulness of...information}                       &  &  &                \\
Start/End information       & \m{4.17}, \sd{1.27} & \m{4.48}, \sd{0.67} & \p{0.754}                 \\
Driving-related             & \m{4.00}, \sd{1.35} & \m{4.17}, \sd{1.15} & \p{0.865}                 \\
Pedestrian crossing         & \m{4.33}, \sd{0.98} & \m{4.52}, \sd{0.59} & \p{0.843}                 \\
Route                       & \m{4.00}, \sd{1.41} & \m{4.35}, \sd{0.78} & \p{0.864}                 \\
Unforeseen Events           & \m{4.42}, \sd{1.16} & \m{4.43}, \sd{0.66} & \p{0.502}                 \\
Point of interest           & \m{3.33}, \sd{1.23} & \m{3.87}, \sd{1.10} & \p{0.184}                 \\
Surroundings                & \m{3.50}, \sd{1.57} & \m{3.74}, \sd{1.32} & \p{0.732}                 \\ \hdashline
\textit{Rating of in-vehicle information} \textit{system}                      &  &  &                \\
Light                       & \m{3.50}, \sd{1.24} & \m{4.04}, \sd{0.88} & \p{0.209}                 \\
Dashboard                   & \m{3.58}, \sd{0.79} & \m{4.30}, \sd{0.82} & \p{0.015} \textbf{*}               \\
Sound                       & \m{4.17}, \sd{0.83} & \m{4.17}, \sd{1.03} & \p{0.764}                 \\
Voice                       & \m{4.00}, \sd{1.21} & \m{4.13}, \sd{1.01} & \p{0.824}                 \\ \bottomrule
\end{tabular}
\label{tab:ratings}
\end{table*}

% Open Feedback
Open feedback regarding the concepts was mainly positive, and participants especially appreciated the focus on BVIs.  [P34, BVI] especially highlighted the information provided after reaching the rest area. In addition, there were some suggestions regarding the messages from sighted people and BVIs. 
[P11, sighted] stated that ``The existence of car driving-related information (turning, pedestrian detection) makes me more anxious and feel less safe to drive, because if no sound is played promptly when we humans see the events (such as pedestrians), we may be worried that the car didn't see it or detect it.''
[P27, BVI] highlighted the ``great design'' besides wanting the option to toggle off ``the sightseeing stuff''. [P32, BVI] wished for earlier auditory signals. In the current implementation, these were played 2s before the maneuver. Regarding additional information to be conveyed, [P34, BVI] stated, ``I'm blind, so I need a lot more info, e.g., current pace, percent of distance traveled.'' [P35, BVI] also suggested announcing ``major streets and squares within towns''.

\section{Discussion}
We presented the results of a study with \N{35} participants, of which 12 were BVIs. The study showed that additional auditory feedback improved trust, understanding, and user experience. These improvements align with previous work showing that transparency improves trust~\cite{koo2015did, du2019look}.
Importantly, prior work on transparency in AVs has predominantly operationalized this through visual representations of system state, such as object highlighting or trajectory visualization~\cite{colley2021effects, schneider2021explain, colley2024effects}. Our results extend this line of work by demonstrating that transparency cannot be reduced to visual availability, but must be grounded in modality-appropriate communication that remains interpretable under constrained perception.

However, regarding subjective SA, the omnibus effect was significant but no pairwise post-hoc comparisons survived corrections. This is notable given prior work emphasizing the importance of SA for BVIs through auditory and multimodal feedback~\cite{brewer2020supporting, brinkley_exploring_2020}. One explanation is that while our auditory design improved perceived understanding, standard SA measures such as the SART may not capture this mediated, explanation-driven awareness. This suggests a limitation of current SA operationalizations in accessibility contexts rather than an absence of effect. 
Moreover, our results are consistent with \citet{brinkley_opinions_2017}, in which participants indicated that AVs need a system that provides information to vehicle occupants to increase environmental (i.e., situational) awareness. At the same time, the significant improvement on the Understanding subscale must itself be interpreted cautiously: the SART is known to be sensitive to the sheer amount of information made available to respondents~\cite{endsley1995measurement, salmon2009measuring}, so part of this effect may be inherent to the measure rather than reflecting deeper situational understanding (see \autoref{sec:limitations}).

\subsection{Reaching an Information Saturation}
We developed the information system based on the recommendations of Brewer et al.~\cite{brewer2020supporting, brewer_understanding_2018}, who argued that this information should focus primarily on speech assistants.
Results showed that participants had higher trust in medium and high information. However, the difference between medium and high is negligible.
We found no significant differences in perceived safety. Interestingly, sighted participants reported a higher cognitive load.
Regarding pragmatic quality and the presence of all relevant information, we found a high increase between the low pieces of information and medium information, but no increase with even more information.

We interpret the results to indicate that information saturation is reached with a medium level of information content. Our results show that medium information achieves sufficient interpretability, while additional information introduces redundancy without measurable benefit and increases cognitive cost, thereby extending prior work on information needs of BVIs~\cite{brewer2020supporting, brinkley_exploring_2020} and transparency in AVs~\cite{koo2015did, du2019look} by identifying an effective upper bound of information provision.

In relation to multimodal feedback research~\citet{nadri2021novel}, this indicates that increasing informational richness is not equivalent to increasing effectiveness; rather, effectiveness depends on aligning information content with user goals and temporal relevance. Consequently, AV systems should move beyond fixed information levels toward adaptive information disclosure strategies that prioritize relevance over completeness.

While this aligns with one participant’s view that one should be able to switch POI information on and off, [P34, BVI]’s focus on highlighting the information provided for the final destination counters this argument. This divergence reflects prior findings that BVIs require both continuous environmental awareness and task-specific confirmation (e.g., destination verification)~\cite{brinkley_exploring_2020}. Therefore, future systems should differentiate between persistent navigation-critical information and optional contextual information, enabling dynamic prioritization. Future work should especially focus on the distinction between these two information types.

More fundamentally, the observed divergence in user preferences highlights a structural limitation of fixed interface configurations. Prior work has shown that user needs, particularly for BVIs, vary substantially with respect to information type, timing, and modality~\cite{brinkley_exploring_2020, brewer_understanding_2018}. Designing separate interface variants for all user groups is therefore infeasible in practice due to resource constraints. One promising direction is the use of computational design approaches (e.g., see~\cite{10.1145/3706598.3713514}), which enable the systematic exploration and optimization of information communication design parameters with users in the loop. Such methods can reduce the design burden by identifying user- or subgroup-specific configurations without requiring exhaustive manual design, thereby fostering scalable inclusivity.

\subsection{Heterogeneity Within the BVI Group}
While sighted participants rated the reasonability of the provided information uniformly high, BVI participants' ratings showed substantial variance (\autoref{fig:reasonable}). We see four non-exclusive explanations. First, the BVI group itself was heterogeneous, ranging from moderate visual impairment to blindness. Participants with usable residual vision could cross-check auditory messages against the high-contrast dashboard, rendering parts of the auditory information redundant, whereas blind participants depended entirely on the auditory channel; the same message set can thus appear excessive to one subgroup and indispensable to another. Second, prior experience with speech-based assistive technologies (e.g., screen readers or navigation applications) plausibly shapes expectations regarding verbosity, pacing, and voice characteristics. Third, the qualitative feedback indicates genuinely divergent preferences: [P27, BVI] asked to toggle off the sightseeing information, whereas [P34, BVI] requested \textit{more} information, such as current pace and distance traveled. Fourth, with \textit{n=}12, variance estimates are themselves unstable. Rather than treating this dispersion as noise, we interpret it as converging evidence that no single fixed information configuration can serve the BVI population---underscoring the need for customizability and the computationally supported personalization approaches discussed above.

\subsection{On the Visual Design of the Dashboard}
At first glance, the differences between sighted and BVIs seem trivial. Therefore, there is a danger that one might assume that for BVIs, auditory (or tactile~\cite{meinhardt2024hey}) communication is sufficient, and the visual information design can be focused purely on sighted users. However, there are many gradations of visual impairment or blindness. Therefore, it may still be sensible to use adapted visual systems that meet the guidelines for accessible software~\cite{etsi_normbarrierefrei_nodate}.
This aligns with prior work emphasizing that BVIs often retain partial vision and benefit from appropriately designed visual interfaces~\cite{brinkley_exploring_2020}. At the same time, existing AV research has largely prioritized visualizations optimized for fully sighted users~\cite{colley2020effect, colley2021should}, often resulting in visually dense interfaces. Our findings suggest that these designs introduce unnecessary complexity when equivalent information can be conveyed more effectively through other modalities.

Therefore, the visual design used high-visibility colors and appropriately sized fonts to convey information (see \autoref{fig:unity:dashboard}). Current commercially available implementations seem overloaded with visual stimuli, including animations (e.g., see Mercedes-Benz MBUX\footnote{\url{https://www.mercedes-benz.de/passengercars/technology/mbux.html}, accessed March 31, 2026.}). While we did not compare our design with variations, through the collaboration with an author with a visual impairment, we argue that this design is a good starting point for future work.

In line with work on multimodal interaction in AVs~\cite{nadri2021novel}, our results indicate that effective communication emerges from combining modalities. Auditory explanations can convey temporal and causal structure, while visual displays can provide complementary spatial context when accessible. This supports a design approach based on multimodal redundancy, where critical information is available across channels and can be selectively attended depending on user capability.

\subsection{Practical Implications}
With the advent of AVs, individuals can access private mobility, which could be a breakthrough for greater accessibility, allowing people with various impairments to become more independent of others and of public transport. However, until now, work on AVs has often only scratched the surface when implementing the necessary adjustments for people with disabilities. Manufacturers should see this trend as an opportunity to reach new customer groups. First, however, their designs have to be adapted.
%One possible way forward would be to assess their visual designs with tools such as OpenVisSim~\cite{jones2020seeing}, which allows the simulation of visual impairments. 
Our work provides the first insights into the design of both a visual and an auditory experience. It shows that simple explanations can increase accessibility and user experience for people with and without visual impairments. Nonetheless, users should at least be able to turn off auditory signals.

Accessibility should therefore not be treated as a post-hoc adaptation but as an integral part of the design process, requiring continuous involvement of target users through participatory or co-design approaches.

Finally, consistent with broader accessibility research, the primary barrier to BVIs seems not to be the automation capability itself, but the lack of appropriate interaction mechanisms. Addressing this gap is essential not only for accessibility but also for trust calibration and adoption across all user groups.

\subsection{Limitations}\label{sec:limitations}
First, the sample was moderate in size (\N{35}) and unbalanced (12 BVI vs.\ 23 sighted). Our a priori power analysis targeted the within-subjects and interaction effects of the mixed design; a sensitivity analysis with the realized group sizes shows that these effects were powered to detect medium effects (f $\approx$ 0.22), whereas the between-subjects comparison of BVI and sighted participants was powered only for large effects (f $\approx$ 0.42). Between-group main effects---particularly non-significant ones---should therefore be considered exploratory. Moreover, the BVI group was heterogeneous in residual vision, ranging from moderate impairment to blindness, and the design neither stratified nor statistically adjusted for visual acuity; observed group effects may partly reflect this within-group variability.

Second, the SART, as a subjective SA measure, is known to be sensitive to the volume of information made available to respondents~\cite{endsley1995measurement, salmon2009measuring}. Because our independent variable directly manipulated information volume, higher SA scores under richer \infoContent are, in part, built into the measure---especially for BVI participants, who had no alternative (visual) access to the same information. The SA results should therefore be read as a manipulation-consistent pattern rather than as independent evidence of improved awareness; future work should employ objective, query-based SA measures adapted to accessibility contexts.

Third, the video-based online setting limits ecological validity, particularly for trust, perceived safety, and cognitive load, which participants formed without exposure to real risk, vection, or vestibular cues. While the scenario was designed realistically and exposure time was controlled, judgments about safety-critical events made from a video may not transfer to real rides. Future studies should employ virtual reality for higher immersion and motion-based simulators (e.g.,~\cite{colley2021swivr, 10.1145/3543174.3545252}), and, ultimately, real vehicles.

Fourth, we relied exclusively on subjective self-report measures, which may have introduced common-method bias.

Finally, we varied the \textit{amount} of information while holding its quality, level of detail, and delivery rate constant; varying these dimensions, as well as the timing of announcements and voice characteristics, may reveal further differences between BVI and sighted users. In any case, customizability of the information provision should be available to BVI users.

\section{Conclusion}
Overall, this work presented the visual and auditory design of an AV intended for usage by BVIs. By collaborating with a person with a visual impairment, we could repeatedly assess both designs (visual and auditory) in the implementation phase. In an online video-based study with \N{35} participants (n=12 with a visual impairment), we compared three levels of auditory communication: low, medium, and high  information. This information included start and stop information, crossing pedestrians, announcements of construction and accident sites, POIs, and a description of the surroundings at the final destination. Results showed the necessity of providing auditory information and that both sighted participants and BVIs benefit from this. The results help in making AVs more accessible.

\section*{Open Science}
All sounds, data, and analysis code are available via \url{https://github.com/M-Colley/bvi-auditory-hav}.

\begin{acks}
The authors thank all study participants.
This work was conducted within the project 'SituWare' funded by the Federal Ministry for Economic Affairs and Energy (BMWi).
\end{acks}

\appendix

\bibliographystyle{ACM-Reference-Format}
\bibliography{sample.bib}

\end{document}